# CONTRIBUTORS' PREFERENCE IN OPEN SOURCE SOFTWARE USABILITY: AN EMPIRICAL STUDY

Arif Raza and Luiz Fernando Capretz

Department of Electrical & Computer Engineering, University of Western Ontario, London, Ontario, Canada N6A 5B9

araza7@uwo.ca, lcapretz@eng.uwo.ca

*ABSTRACT*

*The fact that the number of users of open source software (OSS) is practically un-limited and that ultimately the software quality is determined by end user's experience, makes the usability an even more critical quality attribute than it is for proprietary software. With the sharp increase in use of open source projects by both individuals and organizations, the level of usability and related issues must be addressed more seriously. The research model of this empirical investigation studies and establishes the relationship between the key usability factors from contributors' perspective and OSS usability. A data set of 78 OSS contributors that includes architects, designers, developers, testers and users from 22 open source projects of varied size has been used to study the research model. The results of this study provide empirical evidence by indicating that the highlighted key factors play a significant role in improving OSS usability.*

## 1. INTRODUCTION

Open Source Software systems provide their users with both free access and the ability to modify the source code [1]. One of the indications of progress and development in OSS is that it has influenced almost every dimension of the software development arena. The most successful examples include the GNU/Linux operating system, the Apache HTTP server, the Mozilla Firefox internet browser, and the MySQL database system. Since OSS systems have neither the physical nor commercial boundaries of proprietary software, users from all over the world can make use of them. On one hand, this is advantageous because as more and more users are able to access any OSS, there are more chances of improvement. On the other hand, quality assurance (QA) and its measurement and the post- release management of OSS projects are some of the areas where closed source proprietary software is superior. Viorres et al. [2] relate the OSS popularity "*to the audience that OSS addressees*". They identify "*the usability of the product, the support of OS communities, the notion of accessibility and software engineering usability*" as four HCI related challenges to OSS community. Laplante et al. [3] observe that still many organizations feel reluctant in using open source software mainly due to "*an inherent distrust of OSS quality*". However according to them, higher quality level may be achieved using OSS as compared to closed proprietary software.

The International Organization for Standardization and The International Electro technical Commission ISO/IEC 9126-1 [4] categorizes software quality attributes into six categories namely functionality, reliability, usability, efficiency, maintainability and portability. In the standard, usability is defined as *"The capability of the software product to be understood, learned, used and attractive to the user, when used under specified conditions"*. Iivari [5] identifies that OSS is no more targeted to developers alone, rather to its users that include "*a growing number of non-technical, non-computer professional users*" and thus their needs and expectations should be addressed.

From the usability point of view in particular, OSS is expected to face a more challenging environment because there are all types of users who have both technical and non-technical backgrounds, and who come from every corner of the world bringing their unique needs, expectations and demands. Bodker et al. [6] see the lack of user friendly products in OSS as a serious threat towards its popularity and adoption and believe that it is mainly because OSS developers do not have full understanding of user situations. Nichols and Twidale [7] identify that usability problems arise when we consider users and developers distinctly in OSS environment. They believe that leaving some exceptions aside, most of the OSS projects lack in usability proficiency.

This research work contributes in understanding the effects of some key usability factors through empirical investigation that they play a vital role in improving OSS usability. A quantitative survey of OSS contributors that includes architects, designers, developers, testers and users of different OSS projects has been conducted and reported here. The survey has been used to analyze the conceptual model and hypotheses of the study. The results provide the evidence that the stated key factors play an important role towards the improvement of OSS usability.

In the next section we are presenting the literature review regarding software quality issues in open source environment that motivated and helped this study in selecting the key factors for the study. In Section-3 the research model and the hypotheses of this study have been presented. The research methodology, data collection process and the experimental setup have been explained in the first part of Section-4, reliability and validity analysis of the measuring instrument in the second and data analysis procedures in its third part. In Section-5, hypotheses testing and the analysis of the results will be presented. It will be followed by the discussion in Section-6 that also includes the limitations of the study. Finally Section-7 concludes the paper.

## 2. LITERATURE REVIEW

**2.1. Open Source Quality Issues – In General**

In studying such OSS quality related issues as quality control, quality assurance techniques, risk assessment, testing or usability, there is one point on which researchers generally agree: that OSS quality related issues are not the same as those for closed proprietary software ([8], [9] & [10]). Yunwen and Kishida [11] stress the need for more interaction between software users and developers and suggest that the link to the success of an OSS project is to develop a collaborative platform between the two communities. They realize that having free access to and the right to modify a source code are not the only differences between closed proprietary software and OSS. Rather "*the fundamental difference is role transformation of the people involved in the project*".

Aberdour [1] observes that, compared to the proprietary software environment, open source projects have a higher potential to develop faster and improve their quality because more people can access them. He points out that OSS projects are not only reviewed by the software development team itself, but also are peer reviewed by "*unbiased*" people having no vested interest in such projects. Hedberg et al. [12] also link the improvement in OSS quality and usability to issues related to "*naïve, non computer professional*" users. According to them, not only are their numbers growing day by day but also their expectations are increasing. They observe that as open source software developers are neither paid nor have any formal authority, the major issue to assure quality is their degree of commitment.

Syed-Mohamad and McBride [13] in their study find out that OSS quality varies because it is mainly dependent on "*community usage and defect reporting*". They find out that the reliability growth profile of OSS is different from that of proprietary software as rapid changes in code structure are made by developers in frequently released versions of their projects.

According to Porter et al. [14], through the combination of automated defect tracking tools and effective users' efforts, debugging of OSS projects can be magnified. This would of course result in improving the software quality assurance as well. Maki-Asiala and Matinlassi [15] consider open source as an opportunity for software organizations which let them produce software at relatively lower cost and speed up the production rate as well. Otte et al. [16] in their study find that open source projects benefit from high rate of user participation, user testing and peer reviews. They appreciate structured defect handling processes, significant use of configuration and bug tracking tools in OSS projects.

According to Bouktif et al. [17], OSS phenomenon suffers from frequent changes, increase in complexity and quality deterioration. Whereas Laplante et al. [3] argue that organizations can enhance their software quality by using OSS as it offers better "*security, ease of evolution, and the common "ilities": maintainability, testability, reliability, understandability, and operability*". Lee et al. [18] in their empirical study find the significant influence of software quality over user satisfaction. They conclude that both the software quality and user satisfaction have significant effect over general use of OSS. Based on their evaluation they recommend "*usefulness, ease of use, and reliability*" as some of the factors that OSS practitioners shall pay attention to for improving OSS quality.

The following section presents the literature review of the key usability factors considered in this research work.

**2.2. Usability Factors: Literature Review of Concepts**

Benson et al. [8] observe that in an OSS environment the "*feedback cycle with real users*" is missing. According to them, the communication gap between the developers and the users, the lack of target users' profiles and the degree of responsibility of the developers are the main challenges to improving OSS quality. Bouktif et al. [17] identify "*lack of automated feedback*" about the system's quality as one of the major weaknesses of open source environment. They also propose "*feedback-driven communication service*" to send feedback to developer after each commitment. Iivari [19] in her empirical study about user participation in an OSS project also acknowledges "*informative, consultative and participative roles for* *users*". Bevan [20] also considers feedback from users as the most popular form to improve software usability; however he believes "*it leaves open the risk of inadequate final usability*". Crowston et al. [9] also recommend the collection of users' feedback by building a survey into the software to measure users' satisfaction levels.

Addressing usability issues at the software architecture level, which is the foundation of the whole building block of software, can save much effort later on. However, to do that it is necessary to understand users' requirements and accurately set priorities while keeping in mind that usability is an important quality attribute. This is as true for the closed software as it is for OSS. Golden et al. [21] identify a common practice of software architects when they "*assume that usability issues that arise during user testing can be handled with localized modifications*". They observe that this practice costs high when usability related problems occur at the time of testing and this may require redesigning and in the worst case scenarios even re-architecting of the entire system. Viorres et al. [2] believe that the involvement of end users during design and development support rest of the challenges in software lifecycle. They advocate the need of applying HCI principles in the design processes of OSS to make use of their full potential. Bevan [22] also supports the devotion of human centered design resources to earlier stages of software life cycle.

Zhao and Elbaum [23] notice in their survey that OSS quality techniques differ from traditional software practices. They conclude that, unlike systematic activities in traditional software development, OSS quality assurance is dependent on "*revisions, enhancements and corrections*" by actively involved users in the projects. Hedberg et al. [12] believe that although multiple meanings have been attached to user centered design (UCD) methodology, all of them "*emphasize the importance of understanding the user, his/her tasks or work practices and the context of use*". Folmer and Bosch [24] stress three aspects - usability testing that requires user feed-back on a typical system's task; usability surveys that gather usability experts' or software developers' opinions about whether or not a user interface complies with standard usability norms and usability evaluation questionnaires.

As Çetin and Göktürk [25] state, "*one can't improve what is not measured*", and usability aspects cannot be improved in OSS unless there are ways to measure them quantitatively. While they agree that since they are a non-functional quality attribute and that usability and end user requirements are "*subjective*" matters and cannot be measured directly, OSS developers need to recognize the usability level of their projects. Although traditionally the testing of software consumes considerable time, there is a limited amount of formal testing conducted by OSS developers. Aberdour [1] contrasts the "*formal and structured testing*" which is typical in closed software development, with the "*unstructured and informal testing*" in OSS development. Hedberg et al. [12] also point out that "*test coverage, test-driven development and testing performed by developers*" requires more attention in OSS projects through formal and sufficient test plans to ensure the catching of bugs before the release of the software.

Overall, the lack of OSS documentation, which follows the formal life cycle of software development, may be due, in part, to the fact that developers are solely focused on software development and understand their code so well that they do not feel a need for formal documentation. Nevertheless, formal documentation is of great assistance to new users who wish to understand and adapt to a particular system. Aberdour [1] also highlights the lack of documentation for OSS project compared to the extensive documentation in closed proprietary software and stresses the need for complete documentation in OSS projects. The number of users of an OSS project may be taken as being one indicator of its success and popularity [9]. Among other indicators of OSS quality and level of success, they refer to code and documentation quality, user ratings, downloads, and reuse of code.

Nichols and Twidale [7], while studying usability practices in OSS, highlight the need to seriously address usability issues. They refer to the failure of certain commercial closed source software projects with unusable systems or the poor handling of usability issues as indicating that usability was still an "*unresolved*" issue even with the proprietary software, which is more mature and equipped with more resources, both in terms of experienced manpower and financial resources. According to Pemberton [26], since programmers are generally intuitive and differ from the common user, when they develop software they are normally contented with its usability and interface. Referring to the problems in an OSS environment, he states: "*The general public will have an itch they can't scratch; the programmers won't have that itch, and so won't scratch it*".

## 3.    RESEARCH MODEL AND THE HYPOTHESES

In this study we present a research model to analyze the relationship between the key usability factors and the open source software usability. This work empirically investigates the association between these key usability factors and the OSS Usability. The theoretical model to be empirically tested in this study is shown in Fig. 1. We will examine the relationships of five independent variables on OSS usability, which is the dependent variable in this model. Our aim is to investigate the answer to the following research question:

**Research Question**: *How OSS usability can be improved from contributors' perspective?*

There are five independent and one dependent variable in this research model. The five independent variables are called "Usability Factors" in the rest of this paper. They include Users' Feedback, Usability at architectural level, Design Techniques, Usability Assessment and Documentation. The dependent variable of this study is the OSS usability. The multiple linear regression equation of the model is as follows:

$$\text{OSS Usability Improvement} = f_0 + f_1v_1 + f_2v_2 + f_3v_3 + f_4v_4 + f_5v_5 \qquad (1)$$

where $f_0, f_1, f_2, f_3, f_4$ and $f_5$ are the coefficients and $v_1, v_2, v_3, v_4$ and $v_5$ are the five independent variables.

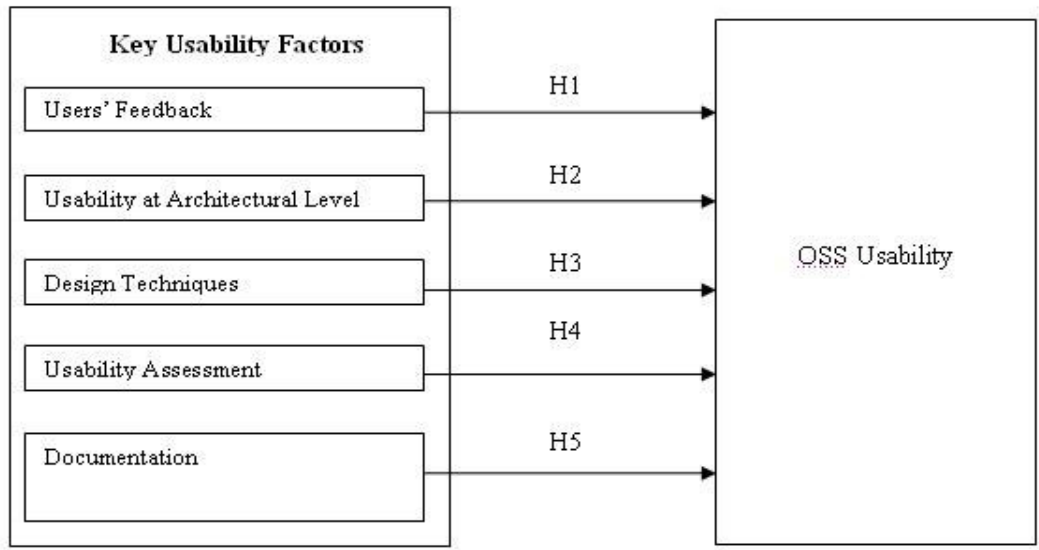

Fig. 1 Research Model

In order to empirically investigate the research question we hypothesize the following:

**H1:** Formal feedback by the users has a positive impact on usability in OSS.

**H2:** Addressing usability issues at software architectural level by the software designers is positively related with improving usability in OSS.

**H3:** User centered design techniques by OSS designers is positively related with improving software.

**H4:** Usability assessment and testing in software have a positive impact on OSS usability.

**H5:** Formal software documentation plays a positive role in improving OSS usability.

## 4. RESEARCH METHODOLOGY

Open source software projects deal with different categories of applications like Database, Desktop environment, Education, Financial, Games / Entertainment, Networking and so on. We sent personalized emails to OSS contributors of different projects. The projects differed in size and range from small to large-scale. We sent our questionnaire to the contributors of projects in the categories of Communications (650), Database (1031), Desktop Environment (136), Education (697), Formats and Protocols (157), Software Development (961), Financial (341) and Games / Entertainment (115) as shown in Fig 2.

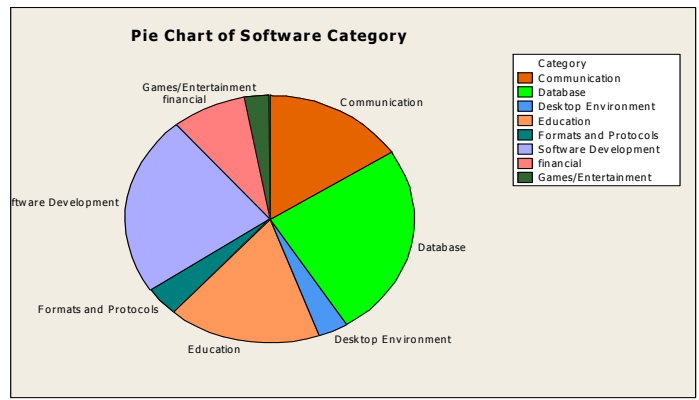

Fig. 2 – Participants' Distribution

We assured the participants that our survey neither required their identity nor would it be recorded. However to support our analysis of data in terms of role of the contributors and their participation in their OSS project, we asked them to share with us their category such as architect/designer, developer, tester or user. This question was optional for the participants to respond unlike the questions related to OSS usability which were mandatory to respond in the survey. Out of 78 respondents altogether, 11 were architects/designers, 32 were developers, 2 categorized themselves as testers and 33 as users.

### 4.1. Data Collection and the Measuring Instrument

The questionnaire presented in Appendix A was used to learn the perceived level of OSS usability improvement as well as up to what extent these usability factors were important for the contributors of OSS projects. We used twenty separate items to measure the independent variables and four items to measure contributors' point of view regarding OSS usability. We reviewed previous researches on the subject of OSS usability so that a comprehensive list of measuring factors could be constructed. To measure the extent to which each of these usability factors have been practiced in OSS projects or agreed upon by OSS contributors, we made use of five-point Likert scale. The Likert scale ranged from "Strongly Agree" (1) to "Strongly Dis-agree" (5) for all items associated with each variable. Four items for each independent variable were designed to collect measures on the extent to which the variable is practiced within each project. The items for all five usability factors are labeled sequentially in Appendix A and are numbered 1 through 20. We measured the dependent variable, i.e. OSS Usability on the multi-item, five-point Likert scale too. The items were specifically designed for collecting measures for this variable and are labeled sequentially from 1 through 4 in Appendix A.

**4.2. Reliability and Validity Analysis of Measuring Instrument**

The two integral features of any empirical study are reliability that refers to the consistency of the measurement, and the validity that is the strength of the inference between the true value and the value of a measurement. For this empirical investigation, we used the most commonly used approaches in empirical studies to conduct reliability and validity analysis of the measuring instruments. The reliability of the multiple-item measurement scales of the five usability factors was evaluated by using internal-consistency analysis, which was performed using coefficient alpha [27]. In our analysis, the coefficient alpha ranges from 0.64 to 0.74 as shown in Table 1. van de Ven and Ferry [28] state that a reliability coefficient of 0.55 or higher is satisfactory, and Osterhof [29] suggests that 0.60 or higher is satisfactory. Therefore we concluded that the variable items developed for this empirical investigation were reliable.

Table 1: Coefficient alpha and Principal Component Analysis (PCA) of variables.

| Usability Factors | Item no. | Coefficient α | PCA Eigen value |
|---|---|---|---|
| Users' feedback | 1 - 4 | 0.65 | 1.48 |
| Usability at architectural level | 5 - 8 | 0.74 | 1.21 |
| Design Techniques | 9 - 12 | 0.64 | 1.48 |
| Usability assessment | 13 - 16 | 0.65 | 1.36 |
| Documentation | 17 - 20 | 0.68 | 1.51 |

According to Campbell and Fiske [30], convergent validity occurs when the scale items are correlated and moves in the same direction in a given assembly. Principal Component Analysis (PCA) [31] was performed for all five key usability factors and reported in Table 1. We used Eigen value [32] as a reference point to observe the construct validity using principal component analysis. In this study, we used Eigen value-one-criterion, also known as Kaiser Criterion ([33] & [34]), which means any component having an Eigen value greater than one was retained. Eigen value analysis revealed that all the five variables completely formed a single factor. Therefore we concluded that the convergent validity can be regarded as sufficient.

### 4.3. Data Analysis Procedure

We analyzed the research model and the significance of hypotheses H1-H5 through different statistical techniques in three phases. Due to the relatively small sample size, both parametric as well as non-parametric statistical approaches were used to reduce the threats to external validity. As our measuring instrument had multiple items for all the five independent variables as well as the dependent variable (refer to Appendix A), their ratings by the respondents were summed up to get a composite value for each of them. In phase-I, tests were conducted for the hypotheses H1-H5 using parametric statistics by determining the Pearson correlation coefficient. For non-parametric statistics, tests were conducted for the hypotheses H1-H5 by determining the Spearman correlation coefficient in phase II. To deal with the limitations of the relatively small sample size and to increase the reliability of the results, the hypotheses H1-H5 of the research model were tested using Partial Least Square (PLS) technique in Phase-III. According to Fornell and Bookstein [35] and Joreskog and Wold [36], the PLS technique is helpful in dealing with issues such as complexity, non-normal distribution, low theoretical information, and small sample size. The statistical calculations were performed using minitab- 15.

## 5.  HYPOTHESES TESTING AND RESULTS

### 5.1. Phase-I

To test the hypotheses H1-H5 of the research model (shown above in Fig. 1), parametric statistics was used in this phase by examining the Pearson correlation coefficient between individual independent variables (key usability factors) and the dependent variable (OSS usability improvement). The results of the statistical calculations for the Pearson correlation coefficient are displayed in Table 2. The Pearson correlation coefficient between users' feedback and OSS usability improvement was found positive (0.479) at P < 0.05, and hence justified the hypothesis H1. The Pearson correlation coefficient of 0.212 was observed at P = 0.062 between usability at architectural level and OSS usability improvement and hence found insignificant at P < 0.05. Therefore the hypothesis H2 that deals with usability at architectural level and OSS usability improvement was rejected. The hypothesis H3 was accepted based on the Pearson correlation coefficient (0.481) at P < 0.05, between the usability design techniques and OSS usability improvement. The positive correlation coefficient of 0.361 at P < 0.05 was also observed between the OSS usability improvement and usability assessment which meant that H4 was accepted. Hypothesis H5 was found significant too and thus accepted after analyzing the Pearson correlation coefficient of 0.508 at P < 0.05 between documentation and OSS usability improvement. Hence, as observed and reported above the hypotheses H1, H3, H4, and H5 were found statistically significant and were accepted whereas H2 was not supported and was therefore rejected.

## 5.2. Phase II

Non-parametric statistical testing was conducted in this phase by examining Spearman correlation coefficient between individual independent variables (key usability factors) and the dependent variable (OSS usability improvement). The results of the statistical calculations for the Spearman correlation coefficient are also displayed in Table 2. The Spearman correlation coefficient between users' feedback and OSS usability improvement was found positive (0.428) at P < 0.05, and hence justified the hypothesis H1. For hypothesis H2, the Spearman correlation coefficient of 0.291 was observed with P=0.01; hence at P < 0.05 significant relationship was found between usability at architectural level and OSS usability improvement in this test. The hypothesis H3 was accepted based on the Spearman correlation coefficient (0.477) at P < 0.05, between the design techniques and OSS usability improvement. The positive Spearman correlation coefficient of 0.318 at P < 0.05 was also observed between the OSS usability improvement and usability assessment which meant that H4 was accepted. Hypothesis H5 was found significant too and thus accepted after analyzing the Spearman correlation coefficient of 0.568 at P < 0.05 between documentation and OSS usability improvement. Hence, as observed and presented above all the hypothesesH1, H2, H3, H4 and H5 were found statistically significant and were accepted in the non-parametric analysis.

Table 2: Hypotheses testing using parametric and non-parametric correlation coefficients

| Hypothesis | Usability Factor | Pearson Correlation coefficient | Spearman Correlation coefficient |
|---|---|---|---|
| H1 | User's feedback | 0.479* | 0.428* |
| H2 | Usability at architectural level | 0.212** | 0.291* |
| H3 | Design Techniques | 0.481* | 0.477* |
| H4 | Usability assessment | 0.361* | 0.318* |
| H5 | Documentation | 0.508* | 0.568* |

\* Significant at P < 0.05. \*\* Insignificant at P > 0.05.

### 5.3. Phase III

In order to do the cross validation of the results obtained in Phase I and Phase II, Partial Least Square (PLS) technique was used in this phase of hypotheses testing. The direction and significance of hypotheses H1–H5 were examined. In PLS, the dependent variable of our research model i.e. OSS usability was placed as the response variable and independent key usability factors as the predicate. The test results that contain observed values of path coefficient, $R^2$ and F-ratio have been shown in Table 3. The users' feedback was observed to be significant at $P < 0.05$ with path coefficient 0.763, $R^2$: 0.23 and F-ratio as 22.68. Usability at architectural level had path coefficient of 0.382 with $R^2$: 0.045 and F-ratio of 3.59 and found insignificant at $P < 0.05$ (with observed P = 0.062). Usability design techniques were observed to have the same direction as proposed in the hypothesis H3 with path coefficient: 1.03, $R^2$: 0.23 and F-ratio: 22.86 at $P < 0.05$. Usability assessment was also found in conformance with the hypothesis H4 with observed values of path coefficient: 0.522, $R^2$: 0.13 and F-ratio: 11.39 at $P < 0.05$. And finally documentation (path coefficient: 1.08, $R^2$: 0.258 and F-ratio: 26.44 at $P < 0.05$) was also found in accordance with H5. Hence in this phase, like in phase I, the hypothesis H2 that deals with usability at architectural level and OSS usability improvement was not found to be statistically significant at $P < 0.05$.

Table 3: Hypotheses testing using Partial Least Square (PLS) regression

| Hypothesis | Usability Factor | Path Coefficient | $R^2$ | F- Ratio |
|---|---|---|---|---|
| H1 | User's feedback | 0.763 | 0.23 | 22.68* |
| H2 | Usability at architectural level | 0.382 | 0.045 | 3.59** |
| H3 | Design Techniques | 1.03 | 0.23 | 22.86* |
| H4 | Usability assessment | 0.522 | 0.13 | 11.39* |
| H5 | Documentation | 1.08 | 0.258 | 26.44* |

* Significant at P < 0.05.  ** Insignificant at P > 0.05

## 5.4. Testing of the Research Model

The multiple linear regression equation of our research model is depicted by Equation-1. The purpose of research model testing was to provide empirical evidence that our key factors play a significant role in improving open source software usability. The testing process consists of conducting regression analysis and reporting the values of the model coefficients and their direction of association. We placed OSS usability as response variable and key factors as predicators. Table 4 displays the regression analysis results of the research model. The path coefficient of four out of five variables: users' feedback, design techniques, usability assessment and documentation were found positive and their t-statistics was also observed statistically significant at P < 0.05. The path coefficient of usability at architectural level was found negative. Negative t-statistics and P > 0.05 (P=0.094) make usability at architectural level statistically insignificant in this research model. $R^2$ and adjusted $R^2$ of overall research model were observed as 0.501and 0.465with F-ratio of 14.24 significant at P < 0.05.

Table 4: Multiple Linear Regression Analysis of the research model.

| Model coefficient Name | Model coefficient | Coefficient value | t-value |
|---|---|---|---|
| Users' feedback | $f_1$ | 0.316 | 3.13* |
| Usability at architectural level | $f_2$ | -0.164 | -1.69** |
| Design Techniques | $f_3$ | 0.172 | 1.51* |
| Usability assessment | $f_4$ | 0.194 | 2.16* |
| Documentation | $f_5$ | 0.465 | 5.02* |
| Constant | $f_0$ | 0.59 | 0.10* |

* Significant at P < 0.05.  ** Insignificant at P > 0.05

## 6.    DISCUSSION

The use of free and open source software (OSS) has increased markedly in recent years largely due to the accessibility and availability of the Internet. However, among other challenges to OSS such as geographically distributed developers, minimal documentation and post release software management - the experience of the end user has become an important issue. Although it is generally believed that OSS is popular with technically adept users, that belief created a blurred boundary between developers and users of open source software. Benson et al. [8] feel the need of making suitable usability methodology as a top priority. They stress upon the contribution HCI professionals to make OSS usable and widely accepted. Çetin and Göktürk [25] claim that only through its measurement and analysis high usability of an OSS project can be achieved. Although they have proposed some metrics for usability assessment, their validation has not been provided. Hedberg et al. [12] state that what is required is to understand the user and the context of use as well as active involvement of the target users through their feedback at an earlier stage of the software design.

Because OSS relies heavily on its users' feed- back to improve its quality, bug reporting by the software users is crucial [9]. One of the reasons for the limited degree of bug reporting has been correctly identified by [7]. They observe that an ordinary non-technical user is generally unable to describe the difficulties s/he faces in a graphical user interface (GUI), and as a result s/he refrains from reporting them. 82% respondents of our survey agree that users' feedback is useful if taken in every phase such as during requirements, design, development, pre and post release. Our empirical investigation also supports the hypothesis that feedback by the users has a positive impact on usability in OSS.

Golden et al.  [21] show their concern regarding usability issues not being addressed at software architecture design level. They identify the repercussions in terms of redesigning or re-architecting of whole system due to such common practices. Nakagawa et al. [37] identify the lack of detailed work to explore how software architecture can influence OSS quality. Through a case study they have shown that software architecture is positively associated with the OSS quality. However our parametric analysis, PLS regression and multiple regression analyses do not support the positive relationship between incorporation of usability issues at architectural level and OSS usability improvement.

Cetin and Gokturk [10] speak of the lack of user-centered design and usability problems in an OSS development environment. They feel that there is a need for more collaboration between interaction designers and OSS developers as well as the early contribution of usability experts in an OSS project to ensure its overall quality. However, Iivari and Iivari [38] do not consider user centered design as a "*separate system development approach*" as they believe it neither covers aspects of system development in the requirements phase nor in the technical implementation phase. 79% respondents of the survey we conducted support the opinion that standardized design techniques can act as a checklist against which software may be inspected. In all the phases of our empirical analysis we have found a significant relationship between design techniques and OSS usability improvement.

Crowston et al. [9] emphasize that not only the success but also the measurement of success and quality of an OSS project are necessary because millions of users are dependent on OSS systems. Çetin and Göktürk [25] do not see usability as the main driving force behind OSS development, which they believe is "*the freedom of the movement*" that does not imply usability within software. They feel the need of and propose a measurement method/framework to assess OSS projects which is required for their self evaluation. 69% of the respondents of our survey believe that considering distributed environment and different cultural backgrounds of OSS users, usability assessment is of prime importance. We have determined a significant relationship between OSS usability improvement and usability assessment as well in all the phases of our empirical study.

Nichols et al. [39] believe that for the less technically oriented users within the OSS community, it is a challenge to simply report bugs discovered in software let alone perform the debugging. They refer to some examples where target users had significant problems with software behavior and documentation even though the developers were quite content. Otte et al. [16] identify the lack of design documentation, maintenance problems regarding source code in case the developers leave the project or testing complexity issues related to diverse platforms. 90% of the participants of our survey agreed that proper documentation of OSS projects increases understandability and learn-ability of software. All the phases of our empirical investigation support positive impact of proper documentation of projects over OSS usability improvement.

### 6.1. Limitations of the study & Threats to External Validity

Empirical methods such as surveys, experiments, metrics, case studies and field studies are used to investigate both software engineering processes and products [40]. Empirical investigations are subject to certain limitations which is applicable in this study as well. Threats to external validity are conditions that limit the researcher's ability to generalize the results of his/her experiment to industrial practice [41], which was the case with this study. Specific measures were taken to support external validity, for example, a random sampling technique was used to select the respondent from the population in order to conduct experiments. We retrieved the data from the most active (having activity of 90% and above) and OSS projects from sourceforge.net which has huge amount of projects listed. The increased popularity of empirical methodology in software engineering has also raised concerns on the ethical issues ([42] & [43]). We followed the recommended ethical principles to ensure that the empirical investigation conducted and reported here would not violate any form of recommended experimental ethics. Another aspect of validity is concerned with whether or not the study reports results that correspond to previous findings. First of all is the selection of independent variables in this work. We have used five independent variables to relate with the dependent variable of OSS usability improvement. We realize that there could be other key factors that influence improvement of usability but we kept the scope of this study within open source software as well as OSS contributors' point of view. Some other contributing factors like OSS development culture, lesser resources of OSS projects as compared to resources of closed proprietary software projects developed in big organizations, voluntary involvement of developers in OSS projects etc have not been considered in this study. Another limitation of this study is its relatively small sample size in terms of number of respondents. Although the proposed approach has some potential to threaten external validity, we followed appropriate research procedures by conducting and reporting tests to improve the reliability and validity of the study, and certain measures were also taken to ensure the external validity.

## 6.  Conclusion

Gaining a better understanding of contributors' opinion through empirical investigation, adapting new approaches to OSS designs to improve usability, and quantifying usability metrics are but three of the challenging options.In this study, we empirically investigate the effect of key factors on OSS usability improvement and find answer to the research question stated in this investigation. Results of this empirical investigation exhibit that the stated key factors of our research model assist in improvement of OSS usability. Empirical results of this study strongly support the hypotheses that users' feedback, design techniques, usability assessment and documentation are positively associated with the usability improvement of an OSS project. However we could not find any statistical significance for "usability at architectural level" on OSS usability improvement, in the phases of parametric, PLS and multiple regression analyses. The study conducted and reported here shall enable OSS development teams to better understand the effectiveness of the relationships of the stated key factors and usability improvement of their projects. The OSS developers need to take into consideration multiple key usability factors to improve usability aspect of software in general and their projects in particular. Currently we are working on to develop maturity model to assess the usability of open source software project, this empirical investigation provides us some justification to consider these key factors as measuring instrument.

International Journal of Software Engineering & Applications (IJSEA), Vol.1, No.2, April 2010

**Authors**

Arif Raza received his M.Sc. (1994) in Computing Science from Birkbeck College, University of London (U.K). Mr. Raza has several years of teaching experience in Computer Science and Software Engineering. He has authored and co-authored several research articles in conference proceedings in the area of software engineering. His current research interests include empirical investigation regarding usability improvement and development of maturity model to assess the usability of open source software project. Currently he is pursuing his Ph.D. studies under the supervision of Prof. Luiz F. Capretz at the University of Western Ontario (Canada) and can be reached at araza7@uwo.ca.

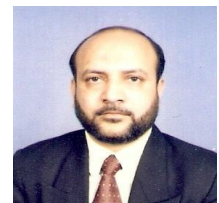

Luiz Fernando Capretz has almost 30 years of international experience in the software engineering field as a practitioner, manager and educator. Having worked in Brazil, Argentina, U.K., Japan, Italy and the UAE, he is currently an Associate Professor and the Director of the Software Engineering Program at the University of Western Ontario, Canada. He has published over 100 peer-reviewed research papers on software engineering in leading international journals and conference proceedings, and he has co-authored two books in the field. His present research interests include software engineering (SE), human factors in SE, software estimation, software product lines, and software engineering education. Dr. Capretz received his Ph.D. in Computing Science from the University of Newcastle upon Tyne (U.K.), his M.Sc. in Applied Computing from the National Institute for Space Research (INPE, Brazil), and his B.Sc. in Computer Science from State University of Campinas (UNICAMP, Brazil). He is an IEEE senior member, ACM member, MBTI certified practitioner, Professional Engineer in Ontario (Canada), and he can be contacted at lcapretz@eng.uwo.ca.

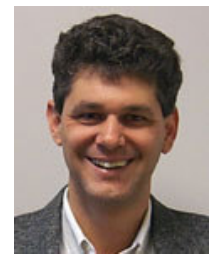

**Appendix A. Key Usability Factors from OSS Contributor's Point of View**

**User's feedback**
1. Users' feedback is useful if taken in every phase such as during requirements, design, development, pre and post release.
2. Quantification of users' feedback and contribution towards an OSS project is must.
3. User's feedback is more effective if recorded with a user's profile.
4. User's feedback is required for application software only.

**Usability at architectural level**
5. Incorporating usability at architectural level saves later modification efforts in terms of time and money.
6. It is impractical to incorporate usability at software architectural level.
7. Usability issues arise at the time of graphical user interface (GUI) development.
8. Usability at architectural level is only an issue of large projects.

**Design Techniques**
9. Standardized design techniques are needed for software usability.
10. Considering different functionality and context of use of software, standardized design techniques for usability are not feasible.
11. There is a need of standardized user interface guidelines for OSS developers.
12. Standardized design techniques can act as a checklist against which software may be inspected.

**Usability assessment**
13. Usability assessment of software is relevant in its context of use only.
14. Considering distributed environment and different cultural backgrounds of OSS users, usability testing is of prime importance.
15. There is no need of formal usability assessment as through developers' peer reviews and community feedback OSS usability is already being evaluated.
16. Usability assessment of an OSS project enhances its quality assurance in general and its reliability and acceptability in particular.

**Documentation**
17. Proper documentation of OSS projects increases understandability and learn-ability of software.
18. Considering voluntary work by OSS developers and frequent release of versions, documentation of OSS projects is impracticable.
19. OSS with proper documentation becomes a better alternative to proprietary software.
20. Proper documentation increases OSS acceptability to general users as well as to organizations.

**OSS Usability**
1. Since OSS quality relies heavily on users' participation and peer reviews, usability issues need to be addressed more seriously.
2. OSS usability cannot be improved unless properly assessed and measured.
3. Success of software depends equally on its correct functionality and usability.
4. Usable software attracts users such as elderly, children as well as people having some disability.